\documentclass[prb,twocolumn,floatfix,superscriptaddress,amsmath,citeautoscript,aps,longbibliography]{revtex4-2}
\pdfoutput=1
\usepackage[T1]{fontenc}\usepackage[latin1]{inputenc}
\usepackage{dcolumn,bm,graphicx,color,booktabs,microtype,afterpage}
\graphicspath{{Figures/}, {FiguresAppe/}}
\renewcommand{\figurename}{Fig.}
\renewcommand{\tablename}{Table}
\makeatletter\renewcommand{\fnum@figure}[1]{\figurename~\thefigure. }\makeatother
\makeatletter\renewcommand{\fnum@table}[1]{\tablename~\thetable. }\makeatother
\newcount\hh \newcount\mm
\hh=\time \divide\hh by 60
\mm=\hh \multiply\mm by 60 \mm=-\mm
\advance\mm by \time
\def\now{\number\hh:\ifnum\mm<10{}0\fi\number\mm}
\usepackage[colorlinks,plainpages=false,linkcolor=blue,urlcolor=blue,citecolor=blue,pdfpagemode=UseNone,pdfstartview=FitBH]{hyperref}

\def\sfo{SrFeO$_\text{3}$}

\def\three{Sr$_\text{3}$Fe$_\text{2}$O$_\text{7}$}

\begin{document}

\title{Anomalous quasielastic scattering contribution in the centrosymmetric multi-$\mathbf{q}$ helimagnet SrFeO$_3$}

\author{Nikita D.\ Andriushin}
\affiliation{Institut f\"ur Festk\"orper- und Materialphysik, Technische Universit\"at Dresden, 01069 Dresden, Germany}

\author{Justus Grumbach}
\affiliation{Institut f\"ur Festk\"orper- und Materialphysik, Technische Universit\"at Dresden, 01069 Dresden, Germany}

\author{Anton A.\ Kulbakov}
\affiliation{Institut f\"ur Festk\"orper- und Materialphysik, Technische Universit\"at Dresden, 01069 Dresden, Germany}
\author{Yuliia V. Tymoshenko}
\affiliation{J\"ulich Center for Neutron Science at MLZ, Forschungszentrum J\"ulich GmbH, 85748 Garching, Germany}
\affiliation{Institut f\"ur Festk\"orper- und Materialphysik, Technische Universit\"at Dresden, 01069 Dresden, Germany}
\author{Yevhen A. Onykiienko}
\affiliation{Institut f\"ur Festk\"orper- und Materialphysik, Technische Universit\"at Dresden, 01069 Dresden, Germany}

\author{Reza Firouzmandi}
\affiliation{Institut f\"ur Festk\"orperforschung, Leibniz IFW Dresden, 01069 Dresden, Germany}
\author{Erjian Cheng}
\altaffiliation[Current affiliation:~]{Max Planck Institute for Chemical Physics of Solids, 01187 Dresden, Germany}
\affiliation{Institut f\"ur Festk\"orperforschung, Leibniz IFW Dresden, 01069 Dresden, Germany}

\author{Sergey Granovsky}
\affiliation{Institut f\"ur Festk\"orper- und Materialphysik, Technische Universit\"at Dresden, 01069 Dresden, Germany}

\author{Yurii Skourski}
\affiliation{Dresden High Magnetic Field Laboratory (HLD-EMFL), Helmholtz-Zentrum Dresden-Rossendorf, 01328 Dresden, Germany}

\author{Jacques Ollivier}
\affiliation{Institut Laue-Langevin, 71 Avenue des Martyrs, CS 20156, 38042 Grenoble CEDEX 9, France}

\author{Helen C.\ Walker}
\affiliation{ISIS Neutron and Muon Source, Rutherford Appleton Laboratory, Chilton, Didcot OX11 0QX, United Kingdom}

\author{Vilmos Kocsis}
\affiliation{Institut f\"ur Festk\"orperforschung, Leibniz IFW Dresden, 01069 Dresden, Germany}
\author{Bernd B{\"u}chner}
\affiliation{Institut f\"ur Festk\"orperforschung, Leibniz IFW Dresden, 01069 Dresden, Germany}
\affiliation{Institut f\"ur Festk\"orper- und Materialphysik, Technische Universit\"at Dresden, 01069 Dresden, Germany}
\affiliation{W\"urzburg-Dresden Cluster of Excellence on Complexity and Topology in Quantum Matter\,---\,ct.qmat, Technische Universit\"at Dresden, 01069 Dresden, Germany}
\affiliation{Center for Transport and Devices, Technische Universit\"at Dresden, 01069 Dresden, Germany}

\author{Bernhard Keimer}
\affiliation{Max-Planck-Institut f\"ur Festk\"orperforschung, 70569 Stuttgart, Germany}

\author{Mathias Doerr}
\affiliation{Institut f\"ur Festk\"orper- und Materialphysik, Technische Universit\"at Dresden, 01069 Dresden, Germany}

\author{Dmytro S.\ Inosov}
\email{dmytro.inosov@tu-dresden.de}
\affiliation{Institut f\"ur Festk\"orper- und Materialphysik, Technische Universit\"at Dresden, 01069 Dresden, Germany}
\affiliation{W\"urzburg-Dresden Cluster of Excellence on Complexity and Topology in Quantum Matter\,---\,ct.qmat, Technische Universit\"at Dresden, 01069 Dresden, Germany}

\author{Darren C.\ Peets}
\email{darren.peets@tu-dresden.de}
\affiliation{Institut f\"ur Festk\"orper- und Materialphysik, Technische Universit\"at Dresden, 01069 Dresden, Germany}
\affiliation{Max-Planck-Institut f\"ur Festk\"orperforschung, 70569 Stuttgart, Germany}

\begin{abstract}\noindent
Centrosymmetric compounds which host three-dimensional topological spin structures comprise a distinct subclass of materials in which multiple-$\mathbf{q}$ magnetic order is stabilized by anisotropy and bond frustration in contrast to the more common path of antisymmetric exchange interactions. Here we investigate static and dynamic magnetic properties of the cubic perovskite SrFeO$_3$\,---\,a rare example of a centrosymmetric material hosting two types of topological spin textures: skyrmion- and hedgehog-lattice phases. Our detailed magnetization and dilatometry measurements describe the domain selection processes and phase transitions in SrFeO$_3$. Spin excitations are investigated using inelastic neutron scattering for all three zero-field phases. In the higher-temperature ordered phases, high-energy magnons increasingly lose coherence, so that spin fluctuations are dominated by a distinct quasielastic component at low energies. We anticipate that this could be generic to symmetric helimagnets in which the chiral symmetry is spontaneously broken by the magnetic order.

\end{abstract}

\maketitle

\section{Introduction}

Magnetic materials which exhibit a superposition of several spiral spin modulations (thereby forming multi-$\mathbf{q}$ spin structures) serve as a natural foundation for the emergence of topologically nontrivial magnetic textures. One of the best-known materials of this kind is MnSi, which forms in the B20 crystal structure and features a skyrmion phase characterized by the formation of a lattice of magnetic whirls described by a combination of three helical modulations~\cite{Muehlbauer2009}. Skyrmions, along with other unconventional spin textures like merons~\cite{Lu_2020, Augustin_2021} and hopfions~\cite{Zheng_2023}, have been the focus of numerous studies due to their potential applications in spintronics, particle-like properties, and associated anomalous transport effects~\cite{Lee_2009,Neubauer_2009,Wang_2022}. In conventional skyrmion materials, the noncollinear magnetic order is driven by the antisymmetric exchange (Dzyaloshinskii-Moriya) interactions~\cite{Dzyaloshinsky1958, Moriya1960, Dzyaloshinskii1964, Dzyaloshinskii1965, Roessler2006, Togawa2016}. The noncentrosymmetric crystal structure may explicitly break chiral symmetry, hence the magnetic order inherits the sense of chirality from the crystal structure.

Contrary to this scenario, so-called Yoshimori-type helimagnets~\cite{Yoshimori1959, Kaplan1959, Villain1959} preserve inversion symmetry at the level of the crystal lattice, such that antisymmetric exchange is forbidden by symmetry. Multi-$\mathbf{q}$ incommensurate magnetic order can nevertheless form in such systems, resulting in a spontaneous breaking of chiral symmetry. For the limited number of known centrosymmetric multi-$\mathbf{q}$ materials, various stabilization mechanisms based on itinerant electrons, anisotropic interactions, and magnetic frustration have been proposed instead~\cite{Mostovoy_2005, Mostovoy_2005_2, Hayami_2021, Wang_2021, Okumura_2022, Hayami_2022}.

Centrosymmetric compounds exhibiting multi-$\mathbf{q}$ magnetic phases remain scarce. Several gadolinium-based crystals have been reported to feature such spin textures, notably triangular-lattice Gd$_2$PdSi$_3$~\cite{Kurumaji2019,Zhang2020}, tetragonal GdRu$_2$Si$_2$~\cite{Khanh_2020,Khan2022,Wood2023}, and breathing-kagome Gd$_3$Ru$_4$Al$_{12}$~\cite{Hirschberger_2019, Hirschberger_2021}. A recent observation of topological transport properties in cubic intermetallics HoCu, ErCu and TmCu plausibly expands the list of centrosymmetric multi-$\mathbf{q}$ systems~\cite{Simeth_2024}. The tetragonal compounds EuAl$_4$ and EuGa$_4$ and their intermediate compound EuGa$_2$Al$_2$ possibly also have topological magnetic phases~\cite{Shang_2021,Zhu_2022,Moya2022,Vibhakar2023}. While Eu and Gd compounds pose problems for neutron-scattering investigations because of their forbiddingly large absorption crosssections, very few transition-metal compounds from this class are known. In particular, two different iron-based materials, cubic \sfo~\cite{Ishiwata2011,Reehuis2012} and tetragonal \three~\cite{Peets2013,Kim2014,Andriushin_2023}, both belonging to the same Ruddlesden-Popper series, have been reported to have complex magnetic phase diagrams including multi-$\mathbf{q}$ phases. Specifically, \three\ exhibits reentrant behavior with two double-$\mathbf{q}$ phases and an intermediate ``spin-cholesteric'' phase~\cite{Spin_cholesteric}. As for cubic \sfo\, it features a low-temperature low-field double-$\mathbf{q}$ skyrmion-lattice phase which gives way to a quadruple-$\mathbf{q}$ three-dimensional ``hedgehog'' lattice at higher temperatures~\cite{Ishiwata2020}. Since Dzyaloshinkii--Moriya interactions are excluded by symmetry in \sfo\ and the magnetism in both materials is highly similar, the observation of topologically nontrivial spin textures in this family has opened a path to a new class of centrosymmetric multi-$\mathbf{q}$ helimagnets which rely on different interactions.

In this work, our attention is directed toward stoichiometric \sfo\ and its multi-$\mathbf{q}$ phases. In the absence of an external field, the compound exhibits chiral magnetic order below 130~K with propagation vector $(\xi\,\xi\,\xi)$ having an incommensurability $\xi$~=~0.13~\cite{Takeda1972,Reehuis2012}. For external field along the [111] direction, at least five distinct magnetic phases were revealed, two of which were identified as multi-$\mathbf{q}$ through single-crystal neutron diffraction~\cite{Ishiwata2020}. Correspondingly, anomalous transport phenomena associated with topological spin texture have been reported~\cite{Ishiwata2011}. Remarkably, symmetry reductions common for perovskite materials were not observed in \sfo, which has a stable cubic crystal structure~\cite{Hodges2000,Schmidt2002}.  In contrast to CaFeO$_3$~\cite{Takano1983}, BaFeO$_3$~\cite{Gallagher1965} and \three~\cite{Kim2021}, \sfo\ does not exhibit charge disproportionation, with Moessbauer spectroscopy finding only one Fe environment down to low temperature~\cite{Gallagher1964,Lebon2004}. \sfo\ is metallic~\cite{MacChesney1965} thanks to mobile holes on the oxygen ligands\,\cite{Takegami2024}, but the remarkably similar magnetism in insulating \three\ argues against a role for itinerant electrons in the mediation of magnetic interactions~\cite{Kim2014}.  \sfo\ is the prototypical example of a centrosymmetric skyrmion- and hedgehog-lattice material, but much remains to be clarified about its magnetic phase diagram and in particular its magnetic excitation spectrum.

Here we comprehensively study the magnetic phase transitions in \sfo\ by means of magnetization, thermal conductivity, and dilatometry measurements, while the spin excitations are probed using inelastic neutron scattering. We find significant spin fluctuations below the ordering temperature observed as quasielastic neutron scattering.

\begin{figure*}[t]
\includegraphics[width=0.99\linewidth]{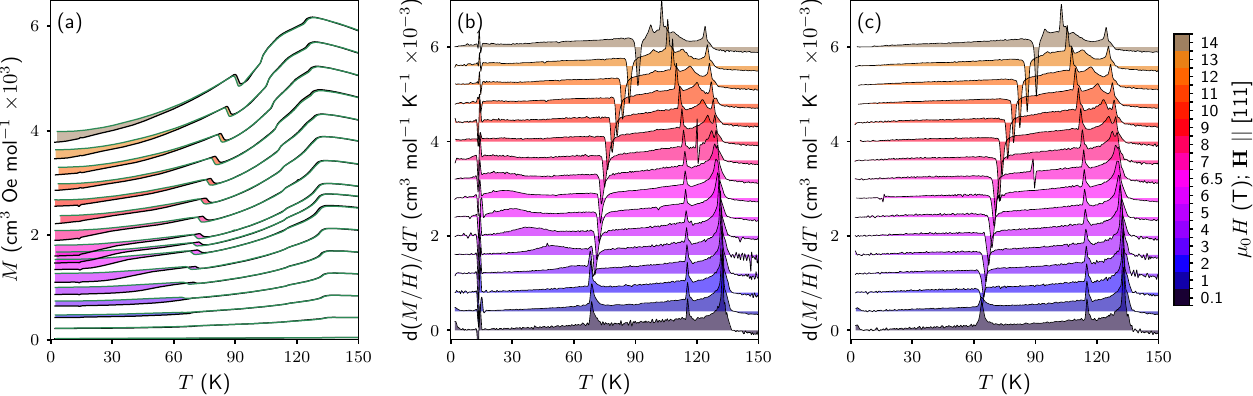}\vspace{3pt}
        \caption{Temperature-dependent magnetization. (a)~$M(T)$ curves at different fields; the shaded area shows the difference between the ZFC (black curves) and FC protocols (green). (b), (c)~The temperature derivatives of magnetization for the ZFC and FC curves, respectively. An offset has been applied for visual clarity. The repeatable anomaly visible at 14~K for the ZFC protocol and identical in all fields arises from a change in cooling mode and is not related to the sample.} \label{MT} 
\end{figure*}

\section{Experimental}

Single crystals of \sfo\ 8~mm in diameter and up to 100~mm in length were grown by the floating zone technique from polycrystalline \sfo\ rods calcined from SrCO$_3$ (99.994\%\ pure, Alfa Aesar) and Fe$_2$O$_3$ (99.998\%\ pure, Alfa Aesar) as described elsewhere~\cite{Peets2012}. To reduce internal cracking, an alumina shield was used to increase temperature gradients and an oxygen partial pressure of 0.8\,atm was used to optimize the crystal's path through the oxygen content phase diagram after growth.  To maximize the oxygen stoichiometry, the resulting crystal was then annealed under 5000~bar of O$_2$ at 500~$^\circ$C for 300 hours, then cooled at 3~$^\circ$C/h to room temperature.

Magnetization measurements for fields $\mathbf{H}\parallel [111]$ were performed by vibrating sample magnetometry (VSM) in a Cryogenic Ltd.\ Cryogen-Free Measurement System (CFMS) using the VSM module, under zero-field-cooled and field-cooled conditions. Four-quadrant $M\!$-$H$ loops were measured at several temperatures. The single crystals were mounted to a plastic rod sample holder using GE varnish. 

Magnetostriction measurements were conducted at the Dresden High Magnetic Field Laboratory at the Helmholtz-Zentrum Dresden-Rossendorf (HZDR), Germany. The optical fiber Bragg grating (FBG) method was employed, wherein the relative change in the sample size was determined through the wavelength shift of reflected light. The measurements were performed using a 60-T pulsed magnet (maximal reached field was 50~T) with a pulse duration of 25~ms and a field rise time of 7~ms.

\begin{figure}
\includegraphics[width=0.99\linewidth]{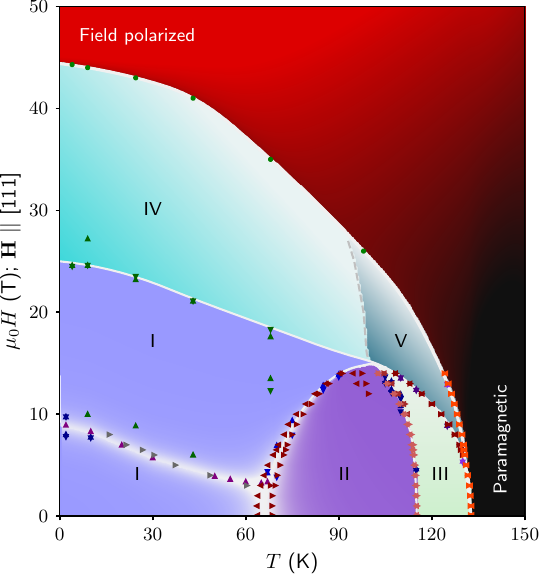}\vspace{3pt}
        \caption{Magnetic phase diagram for the [111] magnetic field direction from magnetization and dilatometry. The orientation of each triangle corresponds to the direction of the temperature or field sweep. Green markers are dilatometry data obtained using a pulse high-field magnet.} \label{PD} 
\end{figure}

For the thermal transport measurements a slab-shaped sample with dimensions 2.44\,$\times$\,0.72\,$\times$\,0.27\,mm$^3$ was mounted in a four-point measurement geometry.
Both the heat current and the magnetic field were applied along the longest edge of the sample, parallel to [111]. 
The heat source was a chip resistor (2.7\,k$\Omega$), while the longitudinal thermal gradient $\nabla_xT=dT/dx$ was measured across a $\Delta x$=1.22\,mm distance using a field-calibrated 3-mil Au/Fe(0.07\%)-chromel thermocouple.
In this experiment, we applied the so-called $\delta$ method, where the longitudinal thermal conductivity $\kappa$ is measured via the difference of the thermocouple signals in the presence and absence of heat current at constant temperature and magnetic field.

The data on low-energy spin-wave excitations were obtained using the cold-neutron time-of-flight spectrometer IN5 at the Institute Laue-Langevin (ILL), Grenoble, France~\cite{Ollivier_2011}. A neutron wavelength of 2.8~\AA\ provided good resolution in low-energy spectra with energy coverage up to 7~meV. Using an orange helium cryostat, the inelastic neutron scattering (INS) data were acquired for the three temperatures 1.5, 100 and 122~K, covering all three low-field magnetic phases. The sample was oriented in the ($HHL$) scattering plane, and the INS data were collected for a region of reciprocal space in the vicinity of the ($1\overline{1}0$) Bragg peak.
 
Complementary INS data with energy transfer up to 43~meV were obtained using the MERLIN time-of-flight spectrometer at the ISIS Neutron and Muon Source, Chilton, Oxfordshire, United Kingdom. The scattering plane was oriented orthogonal to the [111] crystallographic direction. A neutron wavelength of 1.28~\AA\ was used and data were collected in the vicinity of the ($01\overline{1}$) Bragg peak.

\section{Magnetization and phase diagram}

Magnetization measurements on \sfo\ were conducted using an external magnetic field of up to 14~T oriented along the [111] crystallographic direction. Two distinct measurement protocols were employed: the magnetization was measured on warming after cooling in zero field (ZFC) and on cooling in a field (FC). The resulting curves, spanning a broad temperature range, are illustrated in Fig.~\ref{MT}(a), where the shaded area highlights the difference between the two protocols, and the color indicates the magnitude of the external field. The FC protocol overall has a minor excess of magnetization against the ZFC counterpart. The derivatives of magnetization with respect to temperature are also shown to more clearly evidence the various features [Fig.~\ref{MT}(b,c)].

\begin{figure}
\includegraphics[width=\columnwidth]{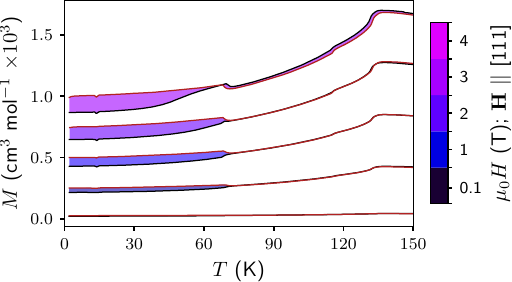}
  \caption{Comparison of temperature-dependent magnetization measured after ZFC (black) and after field training at 8~T (red).}\label{FT} 
\end{figure}

The highest-temperature anomaly is a transition from the paramagnetic state at $T_{\text{N}}~=~$132~K, which exhibits a slight shift to lower temperatures under higher magnetic fields. At a field of 6~T, this transition splits, separating high-temperature ordered phases identified as phases V (in high field) and III (in low field). On cooling, these phases are succeeded by phase II, characterized previously using neutron diffraction~\cite{Ishiwata2020} as a quadruple-$\mathbf{q}$ state with a three-dimensional spin texture in the form of a hedgehog-antihedgehog lattice. Below 66~K, the quadruple-$\mathbf{q}$ order transforms into a double-$\mathbf{q}$ state (phase I), retaining only two propagation vectors from the original set of four. This transition is strongly hysteretic, suggesting it to be first order.  By means of small-angle neutron scattering with polarization analysis, the phase-I spin modulation corresponding to a pair of ordering vectors was determined to comprise a helix and a cycloid, resulting in a magnetic texture breaking the three-fold rotational symmetry of the lattice.

Due to the various combinations of propagation vector pairs within cubic symmetry, phase~I accommodates up to 12 magnetic domains (not counting chiral counterpart domains); this magnetic domain structure leads to the differentiation of the ZFC and FC protocols. The helical component's magnetic moments rotate in the plane orthogonal to the propagation vector. For such order, an external field will destabilize the spin modulation by induction of magnetization only if the field orientation has a nonzero projection on the plane in which the spins rotate. In other words, only fields oriented along the helical propagation vector are compatible with induced magnetization and harmonic helical spin modulation. By this reasoning, domains with helical propagation vectors having maximal projection on the external field vector will be favored over other orientations. In phase I of \sfo, applied fields along [111] exceeding a certain temperature-dependent magnitude can thus reduce the total number of domains from twelve to only three. This irreversible domain selection process is evident in the case of ZFC history as a broad anomaly within phase I in 4--9~T fields [Fig.\ref{MT}(a)]. The phase transitions' critical temperatures and fields were extracted through extremal points in the derivatives and are summarized in the schematic magnetic phase diagram shown in Fig.~\ref{PD}.

Below approximately 4~T, the distinction between ZFC and FC in phase I becomes less apparent due to the inability of a lower field to fully select domains. Here, a third measurement protocol --- field training (FT) --- is more relevant for showing how domain selection influences the low-field magnetization. The field training procedure involves cooling in 8~T to 50~K for domain selection, reducing the field to zero before cooling to base temperature, then subsequently measuring in a lower field on warming. The 8~T field ensures detwinning of the magnetic system from 12 to 3 double-$\mathbf{q}$ domains, and the irreversibility of this process maintains the domain-selected state even in lower fields. The resulting FT magnetization curves, shown in Fig.~\ref{FT}, exhibit an excess magnetization over the ZFC data. However, ZFC curves only serve as a reliable reference for the magnetization of the unselected state to a certain extent. At 4~T, the ZFC curve above 40~K begins to increase towards the FT curve, indicating that at these temperatures, 4~T is sufficient to influence the domain balance in \sfo.

In phase II, the equivalence of all four $\mathbf{q}$ vectors implies the absence of magnetic domains within the quadruple-$\mathbf{q}$ order, and the field history has no effect on the magnetization. Intriguingly, this is also true of phase III.

\begin{figure}
\includegraphics[width=0.99\linewidth]{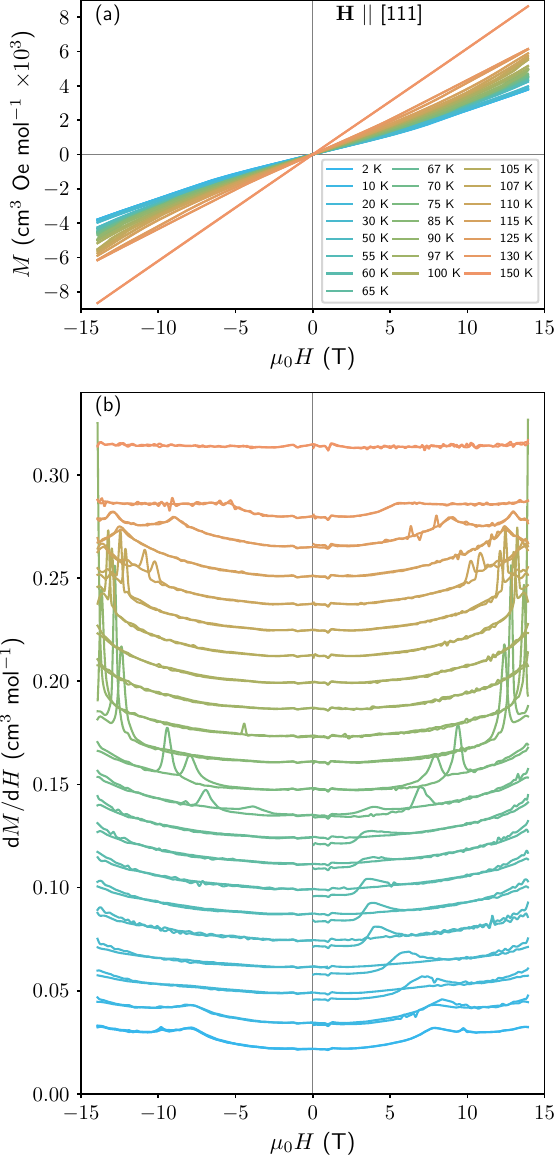}\vspace{3pt}
        \caption{Magnetic field dependence of the magnetization. (a)~$M(H)$ at different fields. (b) The derivatives of the data in panel (a); an offset was applied for visual clarity.  Glitches at $\pm1.0$~T in all datasets are attributed to a range change in the magnet power supply and are not associated with the sample.} \label{MH} 
\end{figure}

Detailed field-dependent magnetization measurements were conducted over a wide temperature range (Fig.~\ref{MH}). Four-quadrant $M(H)$ loops were obtained after cooling in zero field, and include information on the effect of field training as well as phase transitions. Given the strong antiferromagnetic interactions in \sfo, our maximum field of 14~T is far from sufficient to saturate the system into an induced ferromagnetic state (saturation has been reported around 40~T at 4~K~\cite{Ishiwata2011}). Since the phase boundaries among the paramagnetic state, phase III and phase II have relatively weak field dependence in the accessible field range, field sweeps provide limited information on these transitions. At high temperature the main feature is the transition into phase V, visible as a hump in the susceptibility $\text{d}M/\text{d}H$, plotted in Fig.~\ref{MH}(b). In temperatures of 90--100~K above 13~T the magnetization exhibits signs of the transition into high-field phase IV, which is stable above 15~T. However, since 14~T is insufficient to cross this boundary, only the tails of the transition become apparent. In the interval of 65--90~K the transition between phase II and phase I is seen as a peak in the susceptibility. As was seen in the temperature-dependent magnetization, this transition is hysteretic (the transition II~$\rightarrow$~I occurs at slightly higher field than I~$\rightarrow$~II).

Below 66~K, phase I is stable for all accessible fields and no phase transitions are expected. The first quadrant of the measurement loop differs from all subsequent quadrants thanks to domain selection, and a broad peak is observed in the susceptibility in the field range where magnetic dewtinning occurs. The temperature dependence of the peak is in agreement with the $M(T)$ curves and magnetostriction presented below. 

At our lowest temperatures, 2 and 10~K, the appearance of the domain selection transition changes. A broad peak is now observed in all four quadrants, and the magnetization returns to its original value at zero field with no clear training effect. This suggests that at these temperatures the domain walls remain pinned, and the destabilized domains instead transform somehow.  As one possible example, the helical components along disfavoured $\langle 111\rangle$ directions can be decomposed into sine components (i) perpendicular to the applied field along a $\langle 110\rangle$ direction, and (ii) perpendicular to that along a $\langle 121\rangle$ direction, of which the former component may remain stable on its own, leading to a sine+cycloid state in the disfavoured domains. 

We also note that even up to 14~T in the temperature-dependent magnetization in Fig.~\ref{MT}(a), ZFC and FC differ throughout phase I, implying that phase I continues to exhibit field training effects in the 3-domain state, presumably associated with the second component.  We have been discussing these states in terms of a single independent helical component, but the cycloidal component will also respond to field and the second $\mathbf{q}$ vector is neither perpendicular nor parallel to the first. The field evolution of the magnetic order in this system, including field training, should not be simple.

The obtained phase diagram, including high-field transitions from magnetostriction data discussed in the following section, is in good agreement with the one previously constructed based on transport and magnetization data~\cite{Ishiwata2011}. In our data, the transition between phases I and II is offset by approximately 10~K towards higher temperatures in low fields. The transition into phase IV and the saturation field at low temperatures also occur $\sim$5~T higher than reported earlier. These differences may be due to small sample misalignments or slight deviations in oxygen content, which would be expected to strongly alter the magnetic transitions in \sfo.

\section{Magnetostriction}

\begin{figure}
\includegraphics[width=0.99\linewidth]{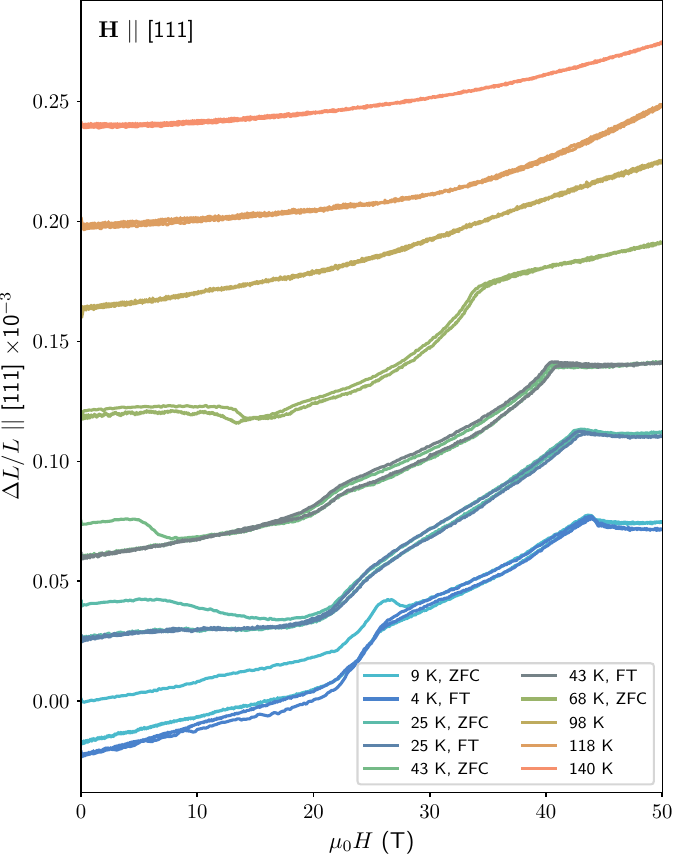}\vspace{3pt}
        \caption{High-field longitudinal magnetostriction (offset applied to the curves for visual clarity). FT stands for the field training protocol.} \label{strict} 
\end{figure}

Due to magnetoelastic coupling, the induced magnetization and formation of magnetic order can result in crystal lattice distortions. These effects are typically relatively weak, though they can be estimated in a dilatometry experiment via highly sensitive measurements of sample linear dimensions. For investigation of magnetostriction effects in \sfo, we utilized the optical fiber Bragg grating method with sensitivity of ${\Delta}L/L\approx10^{-7}$~\cite{Daou_2010}. As a source of magnetic field a 50-T pulsed magnet was used; the sample was oriented for $H\parallel [111]$, and the corresponding change in sample length was probed in the identical direction. Consequently, the longitudinal component of magnetostriction was measured in this experimental setup. Given the domain selection processes in phase I, for temperatures below 60~K we performed the measurement sequence after cooling in zero field, then a second supplemental sequence was performed afterward without heating up.

Magnetostriction in \sfo\ at 4--9~K exhibits mostly linear behavior across distinct phases (Fig.~\ref{strict}). A phase transition is observed both for measurements after ZFC and after field training with a critical field of 24~T at low temperature. This is a transition from phase I into high-field phase IV, which is only stable in fields above $\sim$16~T and hence was not covered in our magnetization data. The second sharp transition at low temperatures (saturation field of 44~T) drives \sfo\ into a spin polarized state. These two transitions remain apparent while their critical fields shift to lower values for 25, 43 and 68~K. At higher temperatures the magnetostriction data do not exhibit sharp phase transitions. At 98~K \sfo\ is expected to transit from phase II into phase V, which is observed as a subtle hump at $\sim$12~T. The transition into the field-induced ferromagnetic state becomes obscured at high temperature and is only visible as a broad slope change around $\sim$28~T.

In phase I, a comparison of ZFC and field training sequences reveals a significant impact of domain selection on lattice distortions. A relative change of linear dimensions up to $\sim$2.5$\times10^{-5}$ is observed, corresponding to a contraction along [111] of the unit cell for the three-domain state, but for the first pulse only. The critical field of domain selection is mainly in agreement with magnetization data. However, the irreversible anomaly observed for the ZFC curves is broadened towards higher fields. This can be attributed to the much faster sweep rate of the pulsed magnetic field used for magnetostriction measurements. The millisecond timescale may already be comparable to the characteristic time of meso- macroscopic domain wall recombination, contributing to the observed broadening of the anomaly.

\section{Thermal transport}

\begin{figure}[t]
  \includegraphics[width=\columnwidth]{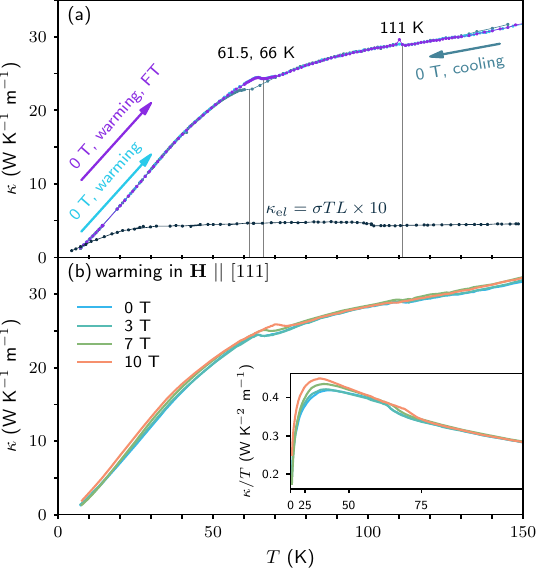}
  \caption{\label{SFO-01}
  (a)~Temperature dependence of the longitudinal thermal conductivity $\kappa$ of \sfo\ in the absence of magnetic field. The black curve corresponds to the $\kappa_{\rm el}=\sigma T L$ electronic component of the heat transport calculated from the Wiedemann-Franz law and the electrical conductivity data of Ref.~\onlinecite{Ishiwata2011}. FT stands for field training protocol.
  (b)~Temperature dependence of $\kappa$ at selected fields $\mathbf{H}\parallel[111]$, shown only for warming. Inset replots the data in panel (b) as $\kappa/T$. The $T$-axis of the inset has a squared scale.}
\end{figure}

Thermal conductivity measurements in the presence of magnetic field can reveal fundamental properties of any heat carriers, such as dimensionality and quantum statistics, distinguish contributions from phonons, electrons, magnons, etc., and identify the mechanisms of their interactions. Figure~\ref{SFO-01}(a) shows the temperature dependence of the longitudinal heat conductivity ($\kappa$) of \sfo\ in the absence of magnetic field. We find that the heat conductivity increases nearly monotonically towards higher temperatures. The contribution of the itinerant charge carriers to $\kappa$ is estimated from the electrical conductivity data of Ref.~\onlinecite{Ishiwata2011} using the Wiedemann-Franz law. We find that a $\kappa_{\rm el}=\sigma T L$ electron contribution (black line multiplied by a factor of 10 for better visibility, $\sigma$ is the electrical conductivity, and $L=2.44\times10^{-8}$~V$^2$K$^{-2}$ is the Lorenz number) is orders of magnitude smaller, and thus negligible compared to the total thermal conductivity. Therefore, the longitudinal heat conduction in \sfo\ is primarily governed by phononic heat transport. The observed monotonic behavior of thermal conductivity, along with the absence of the typical phononic peak below 150~K, cannot be attributed to an increasing number of itinerant electrons. Instead, the formation of short-range magnetic order well above $T_\text{N}$, driven by frustrated magnetic interactions, may contribute into thermal conductivity through scattering phonons on spins.

Similarly to the magnetization measurements, the thermal conductivity measurements also indicate some of the notable phase transitions.
The most notable features are the distinct jumps with broad hysteresis at the boundary between phases I and II, which are observed in the cooling and warming runs at around 61.5 and 66.0\,K, respectively.
We also observe a slight bend around $T$ = 111\,K, the boundary between phases II and III, while we find no anomaly around $T_{\rm N}$ = 132\,K.
In order to compare with magnetostriction and magnetization measurements, we performed a similar field training (FT) sequence; Namely, the sample was cooled in a 5\,T field down to 50\,K, then further cooled to 10\,K in the absence of magnetic field, then we measured on warming in the absence of field.
These field-trained measurements [purple curve in Fig.~\ref{SFO-01}(a)] show no difference to the zero-field cooled measurements (cyan curve), which suggest that the helical domain selection has no effect on the heat transport.

\begin{figure}[t]
  \includegraphics[width=\columnwidth]{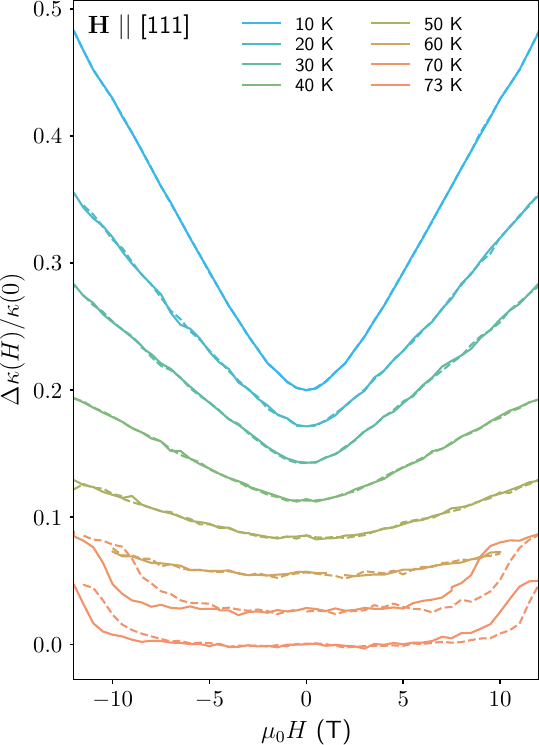}
  \caption{\label{SFO-02}Magnetic field dependence ($\mathbf{H}\parallel{[111]}$) of the relative thermal conductivity change $\Delta\kappa(H)/\kappa(0)$, where $\Delta\kappa(H)$ = $\kappa(H)-\kappa(0)$, for selected temperatures.  The datasets are offset vertically for clarity. Solid and dashed lines correspond to field decrease and field increase directions, respectively.}
\end{figure}

Temperature-dependent thermal conductivity measurements in fields  $\mathbf{H}\parallel[111]$ are shown in Fig.~\ref{SFO-01}(b), while we display the same data as a traditional $\kappa/T$ {\itshape vs}.\ $T^2$ plot as an inset, Fig.~\ref{SFO-01}(c).
In agreement with the magnetization measurements, upon increasing field the boundary between phases I and II shifts towards higher temperatures and the magnitude of the thermal conductivity below $T$ = 66.0\,K is slightly enhanced.
Figure~\ref{SFO-02} shows the field dependence of the relative thermal conductivity change $\Delta\kappa(H)/\kappa(0)$, where $\Delta\kappa(H)$ = $\kappa(H)-\kappa(0)$, for selected temperatures.
In phase II, at around $T$ = 70\,K and in low magnetic fields the $\Delta\kappa(H)/\kappa(0)$ shows a minute and non-monotonic field dependence.
However, as the magnetic field is increased into phase I, $\Delta\kappa(H)/\kappa(0)$ shows a significant, broad, hysteretic jump in accord with the phase boundary in Fig.~\ref{PD}. 
In phase I, the relative thermal conductivity change increases with field and has a symmetric $\sim\sqrt{H^2+H_0^2}$ field dependence with increasing magnitude towards low temperatures, similar to the magnetostriction.
On one hand, this can be explained by the increased gap of the magnon excitations and consequently the decrease in the phonon-magnon scattering~\cite{Buys1982}.
On the other hand, the large magnetostriction can increase the group velocity of the acoustic phonons via an effective hardening of the lattice.
At low temperatures the transition to the 3-domain from the 12-domain region of phase I has no effect on the field dependence of  $\Delta\kappa(H)/\kappa(0)$, in agreement with the field-trained experiments.

\section{Inelastic neutron scattering}

\begin{figure*}[t]
\includegraphics[width=0.99\linewidth]{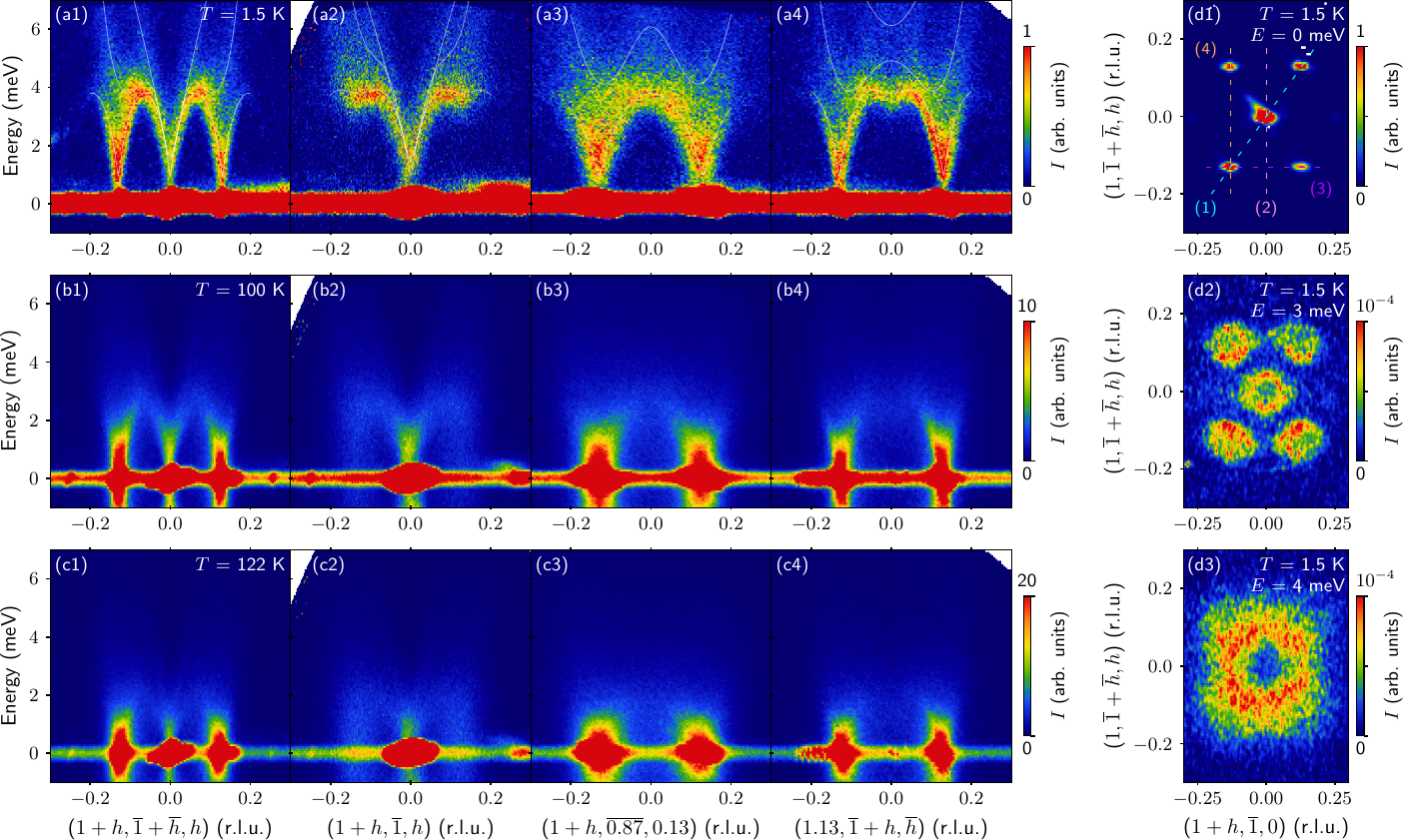}\vspace{3pt} 
        \caption{Low-energy inelastic neutron scattering in vicinity of the $(1\overline{1}0)$ nuclear peak (IN5). Rows correspond to temperatures of 1.5~K (a1--a4), 100~K (b1--b4) and 122~K (c1--c4), and columns correspond to the cuts indicated in panel (d1). The thin white lines are guides to the eye, indicating dispersions of three magnon bands. (d1-d3)~Constant-energy cuts for the ($H\overline{K}K$) reciprocal plane through the ($1\overline{1}0$) nuclear peak at $T = 1.5$~K.}  \label{IN5}
\end{figure*}

\subsection{Low-energy magnons}

\begin{figure}
\includegraphics[width=0.99\linewidth]{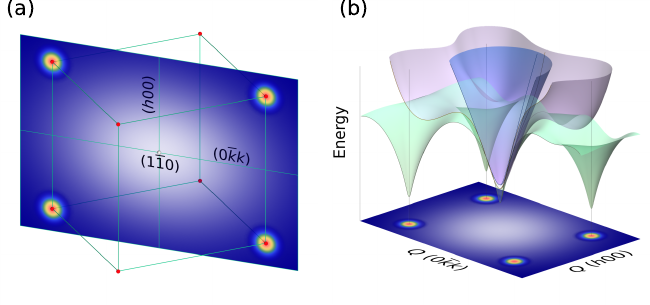}\vspace{3pt}
        \caption{(a)~Schematic representation of the ($H\overline{K}K$) reciprocal-space plane including four magnetic satellites. (b)~The dispersion surfaces of the three low-energy magnon bands for this plane.} \label{Cartoon} 
\end{figure} 

Investigation of the low-energy spin excitations in \sfo\ was carried out with the IN5 time-of-flight neutron spectrometer at the ILL. The measurements were performed at three temperatures corresponding to the three zero-field phases of \sfo, focusing on the magnetic satellites in the vicinity of the ($1\overline{1}0$) nuclear reflection. No external fields were involved, meaning that in case of 1.5~K (phase I) the multidomain state was probed. Given the cubic symmetry, the incommensurate propagation vector $(\xi\,\xi\,\xi)$ ($\xi~=~0.129$ r.l.u.) can lie along any of the 8 $\langle 111\rangle$ directions, and magnetic Bragg peaks from more than one domain can be seen simultaneously via elastic scattering [Fig.~\ref{IN5}(d1)]. The orientation of the ($H\overline{K}K$) scattering plane in reciprocal space is shown schematically in Fig.~\ref{Cartoon}(a).

High-resolution INS measurements revealed the low-energy part of the \sfo\ spectra, which is shown as 2D momentum-energy cuts for different high symmetry directions in reciprocal space [phase I is shown in Fig.~\ref{IN5}(a1-a4), where cuts (1)-(4) are identified in Fig.~\ref{IN5}(d1)]. At 1.5~K, three magnon branches are distinguishable. The most intense mode originating both at the incommensurate ordering vectors and at the nuclear Bragg peak is limited by a $\sim$4~meV bandwidth. Above 4~meV, additional spin excitations with a lower spectral weight, but steeper dispersions, become apparent. The dispersions are marked with thin white lines on top of the measured spectra. Such a spin wave spectrum with several magnon bands most likely stems from the multi-$\mathbf{q}$ magnetic order. For instance, the middle energy band has a non-monotonic dispersion, exhibiting a minimum in the vicinity of the propagation vector of the magnetic structure. The constant-energy cuts demonstrate that the low-energy branch has a well-resolved conical dispersion below 4~meV [Fig.~\ref{IN5}(d2)], while at 4~meV the spectrum is complicated and no longer that well resolved, as the two other branches start gaining spectral weight [Fig.~\ref{IN5}(d3)]. From our full dataset of scattering intensity as a function of momentum and energy transfer, the dispersion surfaces of the three mentioned bands were reconstructed schematically for the ($H\overline{K}K$) reciprocal plane --- see Fig.~\ref{Cartoon}(b). 

\begin{figure}
\includegraphics[width=0.99\linewidth]{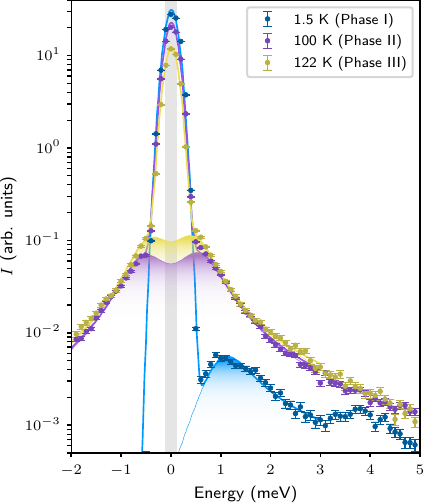}\vspace{3pt}
        \caption{Quasielastic neutron scattering at the propagation vector at three temperatures. The experimental data is the weighted sum of $\mathbf{Q}~=~(0.83, -0.83, -0.13)$ and five more symmetry-equivalent $\mathbf{q}$-points. The shaded curves are Lorentzian contributions after convolution with the resolution function. The grey rectangle indicates the full width at half maximum of the instrumental resolution.} \label{QuasiElastic} 
\end{figure} 

\begin{figure}
\includegraphics[width=0.99\linewidth]{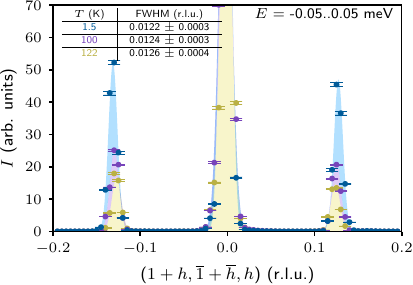}\vspace{3pt}
        \caption{Elastic neutron scattering intensity at three different temperatures. The peak centered at $h = 0$ is the [$1\overline{1}0$] nuclear reflection. The filled areas are Gaussian fits.} \label{Qwidth} 
\end{figure}

The transition into phase II leads to softening of the low-energy spectra in \sfo\ [Fig.~\ref{IN5}(b1-b4)]. The bandwidth of the lowest-energy spin excitations decreases by $\sim$1 meV while overall the dispersions become less resolved. The signal below 1~meV is enriched with additional intensity, which is attributed to spin fluctuations (note the different color scale compared to the 1.5~K colormaps). Due to the increased thermal population of magnons, the inelastic scattering has significant spectral weight in the energy-gain channel ($E < 0$) as well. The transition into phase II also leads to the appearance of new magnon branches below 3~meV in reciprocal directions equivalent to (0$\overline{K}K$), which were not visible in phase I [compare panels (b2) and (a2) in Fig.~\ref{IN5}]. The effect of softening and enhancement of quasielastic scattering intensity is even more prominent in phase III at 122~K [Fig.~\ref{IN5}(c1-c4)]. The remarkable redistribution of intensity from low-energy magnon branches into quasielastic scattering likely stems from frustration of the underlying magnetic interactions. Similar traces of spin fluctuations were observed in related material \three~\cite{Spin_cholesteric}.

\begin{figure*}[t]
\includegraphics[width=0.99\linewidth]{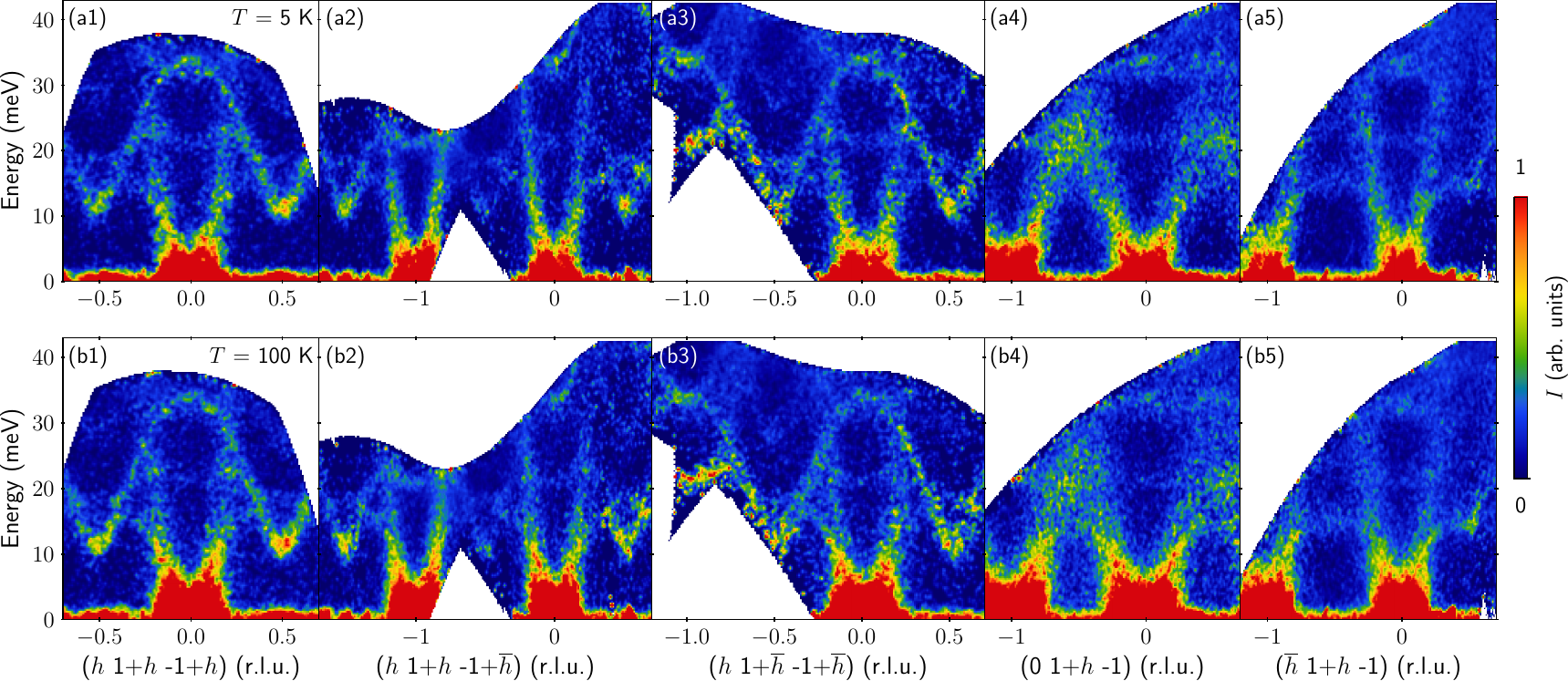}\vspace{3pt}
        \caption{Higher-energy inelastic neutron scattering in the vicinity of the $(1\overline{1}0)$ nuclear peak (MERLIN). Rows correspond to temperatures of 5~K (a1-a5) and 100~K (b1-b5).} \label{MERLIN} 
\end{figure*}

\subsection{Spin fluctuations}

For a more detailed look into the low-energy scattering intensity, we analyze 1D energy cuts at the ordering vectors (Fig.~\ref{QuasiElastic}). For each temperature, the data were integrated in vicinity of the magnetic Bragg peaks and plotted on a logarithmic scale. The experimental intensity curves were taken as a weighted sum of several equivalent $\mathbf{q}$ 
points for improved statistics. The fitting model includes the elastic line, and takes into account the instrumental resolution and background coming from incoherent scattering. The magnetic excitations were modeled by a damped oscillator in the form of a Lorentzian function. The details of the fitting routine are described in Sec.~\ref{app:QENSfit}. The model used fits the experimental data reliably and allows tracking of the low-energy excitation softening. The spin gap is apparent at 1.5~K, with a peak centered at 0.99~meV. However, at higher temperature due to enhancement of spin fluctuations the peak is not visible directly. Fitting with the same model as at low temperature gives values of 0.53 and 0.42~meV at 100 and 122~K, respectively.

The development of broad tails in low energy excitations is evidence of prominent spin fluctuations. The width in energy, beyond the instrumental smearing accounted for during fitting, is directly associated with the temporal coherence of spin excitations and their characteristic lifetime. Similar quasielastic scattering has been observed in frustrated magnets exhibiting spin glass behavior~\cite{Krimmel_2009, Li_2023} and in spin liquids~\cite{Shen_2018, Dai_2021}. However, these examples are notable for their lack of long-range magnetic order, unlike the case of \sfo. Here, such spin fluctuations coexist with the long-range order of phases II and III. This is evident from elastic cuts through the same data (Fig.~\ref{Qwidth}), where we see a clear absence of broadening in the magnetic reflections.

\subsection{High-energy magnons}

The spin excitations above 6~meV in \sfo\ were probed using the MERLIN time-of-flight spectrometer at ISIS. The data were collected in the vicinity of the $(01\overline{1})$ nuclear peak at 5 and 100~K, corresponding to phases I and II, respectively. The momentum-energy cuts along the propagation vector [Fig.~\ref{MERLIN}(a1-a3)] revealed that two magnon branches observed at low energies extend beyond 35~meV. An additional spin-wave mode with a cosine-like dispersion becomes visible at energies between $\sim$10 and $\sim$34~meV. Two other cuts through the $(01\overline{1})$ point [Fig.~\ref{MERLIN}(a4-a5)] show that these excitations are broader here. Modes were similarly broad at low energy. Most likely, not all the modes could be resolved in these data, as a comparison to the multiple magnon branches at higher energy in the IN5 data (Fig.~\ref{IN5}) makes evident --- the two modes originating at the nuclear peak appear as a single branch in the MERLIN data. The high-energy spectra can naturally include phonon excitations, but in \sfo\ the phonon contribution becomes visible only at higher $\mathbf{Q}$ (see details in Sec.~\ref{app:HighQ}). The spectra shown in Fig.~\ref{MERLIN} primarily show magnetic excitations, and the inelastic scattering from the lattice is negligible.

The data at increased temperature, in phase II [Fig.~\ref{MERLIN}(b1-b5)], has only subtle differences compared to low temperature. The energy scale of tens of kelvins is much better seen at low energies, where the temperature can compete with anisotropies and weaker exchange interaction. At higher energies, the modes determined by strong nearest-neighbor interactions dominate, and the dispersions are only weakly affected by temperature.

\section{Discussion and Summary}

The low-energy spectrum of \sfo\ appears to be peculiar: the dispersing multiband excitations are joined in phases II and III by a strong quasi-elastic signal. Heating induces the expected smearing and softening of the magnon bands, but the spin fluctuations becomes significantly more pronounced. The coexistence of such strong magnetic fluctuations with long-range order are intriguing, since conventionally these are observed in states where temperature or frustration destroys long-range order and only short-range correlations are present.

Given the metallic properties of \sfo, the influence of conduction bands on magnon decay could be a factor. However, this is an unlikely scenario, since very similar broadening of spin excitations was observed in the related insulator \three~\cite{Spin_cholesteric}. Both compounds exhibit helical magnetic order and multi-$\mathbf{q}$ phases, and host pronounced spin fluctuations which coexist with well-defined magnon excitations over a wide temperature range.

Both \sfo\ and \three\ possess crystallographic inversion symmetry, which, in conjunction with helical order spontaneously breaking chiral symmetry, leads to energetically equivalent left- and right-handed spin helices. Since the underlying achiral crystal structure does not dictate which chirality should dominate, the system may contain both types of chiral domains and corresponding domain walls. In \sfo, phase II does not exhibit the usual domain selection as a consequence of being a quadruple-$\mathbf{q}$ state, meaning that all magnetic domain walls are chiral domain walls. Phase III should also include such chiral domains; however, details on the magnetic order here are difficult to derive from domain selection, as the absence of domain selection could stem from either a multi-$\mathbf{q}$ state or rapid relaxation at high temperature. In both compounds, it is plausible that the interplay of frustrated exchange interactions, inversion symmetry, and chiral magnetic domains hosted on an achiral lattice is responsible for the intense spin fluctuations. For instance, the chiral domain walls and their associated fluctuations may introduce an additional degree of freedom for spin dynamics. It is also possible that certain components of the modulated spin structure become disordered at high temperature, for instance through transitions from helical order (sine\,+\,$i$sine) to sine (sine\,+\,short-range), giving rise to these fluctuations. In this scenario, while the system exhibits long-range order, not every component of every spin is long-range ordered.

The unconventional magnetic order in \sfo\ is accompanied by multiband low-energy spectra that exhibit intriguing temperature evolution. A thorough theoretical description of the spin excitations will likely require not only modeling of spin waves in the multi-$\mathbf{q}$ state, but also accounting for the thermal fluctuations that are important for phases II and III. Approaches for such modeling are still under development, as the advent of experimental data for centrosymmetric multi-$\mathbf{q}$ materials which rely on bond frustration instead of Dzyaloshinskii-Moriya interactions is still relatively recent. In this regard, \sfo\ serves as a particularly accessible platform for investigating complex multi-$\mathbf{q}$ phases in such systems.

\begin{acknowledgments}
The authors are grateful for experimental assistance from the group of M.\ Jansen, and thank S.\ Nikitin for stimulating discussion. This project was funded by the German Research Foundation (DFG) through individual grants PE~3318/3-1, IN~209/7-1, and IN~209/9-1 (Project Nos.\ 455319354, 285734972, and 434257385); through projects C01, C03, and C07 of the Collaborative Research Center SFB~1143 (Project No.\ 247310070); through the W\"urzburg-Dresden Cluster of Excellence on Complexity and Topology in Quantum Materials\,---\,\textit{ct.qmat} (EXC~2147, Project No.\ 390858490); through the Priority Programme on Skyrmionics SPP~2137 (Project No.\ 360506545); and through the Collaborative Research Center TRR~80 (Project No.\ 107745057). V.K.\ and E.C.\ were supported by the Alexander von Humboldt Foundation. The authors acknowledge the Institut Laue-Langevin, Grenoble (France) for providing neutron beam time~\cite{ILL4011720IN5}. We gratefully acknowledge the Science and Technology Facilities Council (STFC) for access to neutron beamtime at ISIS~\cite{RB1920064Merlin} and for the provision of sample preparation facilities. We acknowledge support of the HLD at HZDR, a member of the European Magnetic Field Laboratory (EMFL).
\end{acknowledgments}

\appendix

\section{Quasielastic neutron scattering fitting} \label{app:QENSfit}

\begin{figure}[b]
\includegraphics[width=0.99\linewidth]{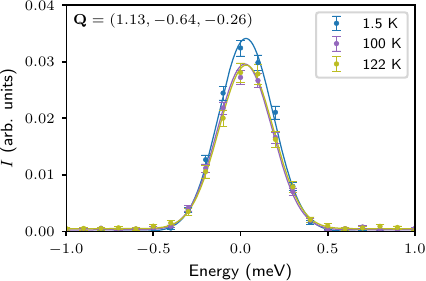}\vspace{3pt}
        \caption{The incoherent neutron scattering at three temperatures. Solid curves are Gaussian fits.} \label{incoherent} 
\end{figure}

To characterize the quasielastic scattering component, it is first essential to eliminate background contributions.  The background of INS spectra at low energies always includes a contribution from incoherent scattering from both sample and sample environment. The incoherent scattering does not depend on the scattering vector and changes only weakly with temperature (following the Debye-Waller factor). It also naturally has a well-defined Gaussian profile in energy transfer. By integrating a region of momentum space containing no elastic signal or low-energy magnons we can extract the intensity corresponding to incoherent scattering. The signal in the vicinity of $\mathbf{Q} = (1 + \xi, -1 + 2\xi, -2\xi)$ with $\xi = 0.13$ is far enough from the main Bragg peaks to be suitable for the estimation of the incoherent contribution (Fig.~\ref{incoherent}). The signal indeed depends weakly on temperature and can be fitted with a Gaussian profile reliably. The absolute values of the intensity of the incoherent scattering are almost three orders of magnitude lower than the magnetic elastic scattering intensity. As this contribution is always the same at all scattering vectors, we use it as an estimate of the background in further quasielastic scattering analysis.

In addition to the background, low-energy spectra in the vicinity of the propagation vector include an elastic peak from scattering on ordered magnetic moments, low-energy magnetic excitations, and the quasielastic scattering peak centered at zero energy. The sum of these three contributions is additionally convoluted with the instrumental resolution function, resulting in a smeared measured signal.

Integration of experimental data can smooth out some features or introduce artificial broadening, therefore we integrate intensity only in the close vicinity of the propagation vector with ranges of $\pm0.015$~r.l.u. in three orthogonal momentum directions. In order to improve the statistics, the final dataset used for analysis was a weighted average of six equivalent $\mathbf{q}$-points.
The general fit model $I(E)$ can be rendered as
\begin{equation} \label{eq:fit}
\begin{split}
    I(E) ={}&\big(S(E) * F_\mathrm{res}(E)\big) + I_\mathrm{incoh}\text{, where}\\
    S(E) ={}&S^{E}\delta(E) + S_{\mathrm{Lor}}(E, E_0)\text{ and}\\
    S_{\mathrm{Lor}}(E, E_0) ={}&\frac{A E}{1 - \mathrm{exp}(-E/k_{\mathrm{B}}T)} \times \\
    &\biggl( \frac{\Gamma}{(E - E_0)^2 + \Gamma^2} +  \frac{\Gamma}{(E + E_0)^2 + \Gamma^2}\biggl).
\end{split}
\end{equation}
Here the neutron scattering intensity $I(E)$ is a convolution of the dynamical spin structure factor $S(E)$ and the instrumental resolution function $F_\mathrm{res}(E)$ given by a Gaussian profile. The contribution from the elastic peak is given by a zero-centered delta function scaled by an amplitude $S^{E}$. The second term $S_{\mathrm{Lor}}$ represents a Lorentzian function centered at $E_0$ with a full width at half maximum $\Gamma$ and amplitude $A$. The Lorentzian peak function includes two terms, corresponding to energy-ain and -loss processes, multiplied by the temperature-dependent Bose factor. The $\Gamma$ parameter is inversely proportional to the characteristic lifetime of the excitation.

The obtained fit is shown in the main text in Fig.~\ref{QuasiElastic}. The fit parameters with calculated confidence intervals are shown in Table~\ref{tab:fit}.

\begin{table}[t]
\caption{The low-energy INS fit results and their uncertainties.} \label{tab:fit}
\begin{tabular}{l|l|c|c|c}
\toprule
$T$~(K) & Parameter & $-\sigma$ & Best & $+\sigma$  \\
\midrule\midrule
1.5~ & $S^{E}$~(arb.~units) & $-0.8$ & 31.3 & 0.8 \\
& $A$~(arb.~units)$\times10^3$ & $-0.07$ & 0.22 & 0.08 \\
& $\Gamma$~(meV) & $-0.2$ & 0.6 & 0.4 \\
& $E_0$~(meV) & $-0.6$ & 1.0 & 0.4 \\
\midrule
100~ & $S^{E}$~(arb.~units) & $-0.4$ & 22.7 & 0.4 \\
& $A$~(arb.~units)$\times10^3$ & $-0.02$ & 0.26 & 0.02 \\
& $\Gamma$~(meV) & $-0.04$ & 0.42 & 0.04 \\
& $E_0$~(meV) & $-0.07$ & 0.56 & 0.06 \\
\midrule
122~ & $S^{E}$~(arb.~units) & $-0.2$ & 12.3 & 0.2 \\
& $A$~(arb.~units)$\times10^3$ & $-0.014$ & 0.281 & 0.015 \\
& $\Gamma$~(meV) & $-0.02$ & 0.37 & 0.02 \\
& $E_0$~(meV) & $-0.04$ & 0.46 & 0.04 \\
\bottomrule
\end{tabular}
\end{table}

\section{Nonmagnetic inelastic neutron scattering} \label{app:HighQ}

\begin{figure}[t]
\includegraphics[width=0.99\linewidth]{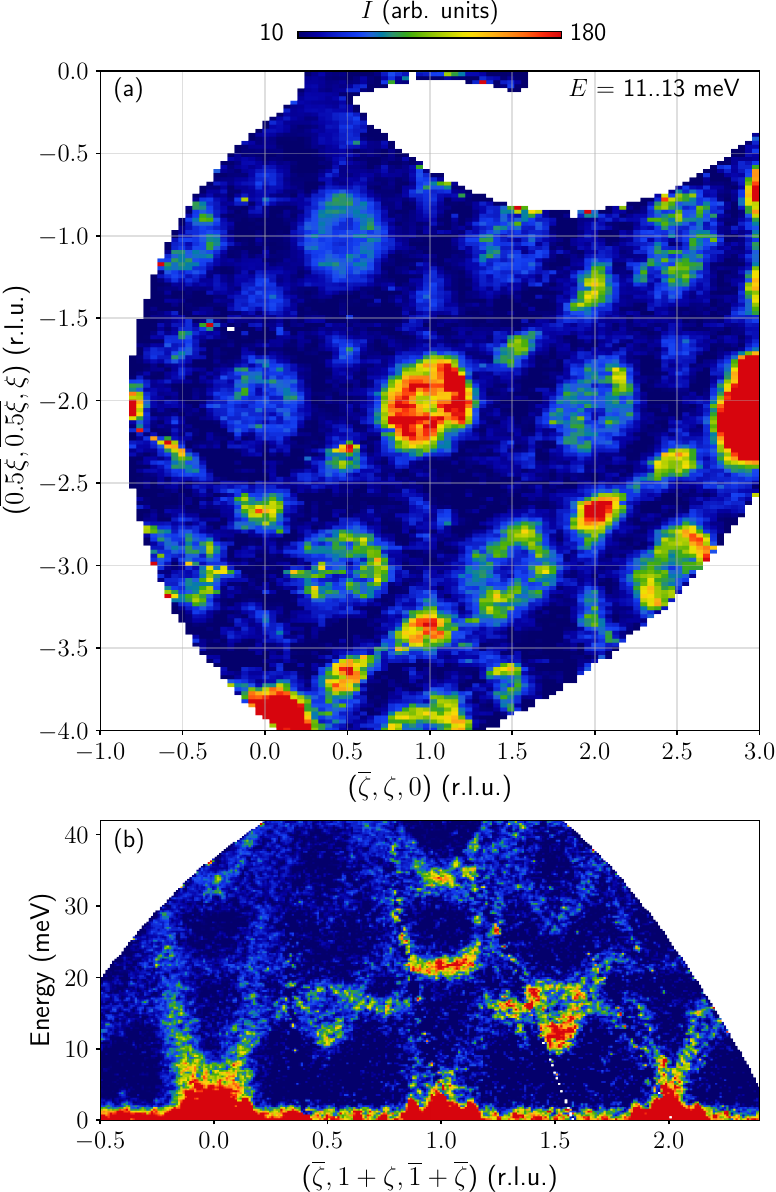}
        \caption{INS data measured at 1.5~K (MERLIN). (a)~Constant energy cut at $12\pm1$~meV for the scattering plane orthogonal to [111]. (b)~Energy-momentum cut along the [$-\zeta, 1 + \zeta, -1 - \zeta$] momentum direction.} \label{Phonons} 
\end{figure}

The scattering plane orthogonal to the [111] direction was measured in the inelastic neutron scattering experiment using the MERLIN time-of-flight spectrometer. The relatively short neutron wavelength (1.28~\AA) allowed us to cover a sufficiently large volume of momentum-energy space. In order to distinguish spin and phonon excitations, we compare the spectra at small and at large $\mathbf{q}$-vectors. The excitations at 12~meV (integrated in the interval 11--13~meV) are shown in Fig.~\ref{Phonons}(a). The spin excitations are most prominently visible at small $\mathbf{q}$ vectors (upper left); the ring at ($0, 1, -1$) [or (0.5, -1) in the figure's coordinates] clearly corresponds to magnon branches. As the wavelength of the scattering vector decreases the magnon spectral weight drops rapidly, while the phonon contribution grows. In the vicinity of more-distant Bragg peaks, the circular-shaped magnons are replaced with a more elliptical signal formed by a combination of several phonons. It is apparent that already at $\mathbf{Q}~=~(0, 2, -2)$ (central red ring) the signal is a mixture of phonon and magnon excitations. The momentum-energy cut in Fig.~\ref{Phonons}(b) also demonstrates that for the [$-\zeta, 1\!+\!\zeta, -1\!-\!\zeta$] momentum direction, the spectra primarily consist of spin excitations for $|\zeta| < 0.5$, while for $|\zeta| > 1.5$ they are dominated by much sharper phonons. This confirms that the broadness of the spin waves is unrelated to the instrumental resolution or sample quality, but is likely due to more than one mode comprising this branch.

\bibliography{bibliography}

\end{document}